\documentclass[aps,pra,reprint,showpacs,superscriptaddress]{revtex4-1}
\usepackage{epsfig}
\usepackage{natbib,color,epstopdf}
\usepackage{amsfonts}
\usepackage{graphicx}
\usepackage{amsmath, amsthm}
\usepackage{bm}
\usepackage{float}
\usepackage{color}
\newcommand{\be}{\begin{equation}}
\newcommand{\ee}{\end{equation}}
\newcommand{\bwt}{\begin{widetext}}
\newcommand{\ewt}{\end{widetext}}
\newcommand{\bea}{\begin{eqnarray}}
\newcommand{\eea}{\end{eqnarray}}
\newcommand{\ket}[1]{|#1\rangle}
\newcommand{\bra}[1]{\langle #1|}

\newcommand{\pro}[1]{\ket{#1}\bra{#1}}

\newcommand{\ip}[2]{\langle #1|#2 \rangle}
\newcommand{\tr}[1]{\mbox{Tr}\left[#1\right]}

\usepackage{lipsum}

\begin{document}
\title{Quantum Control through a Self-Fulfilling Prophecy}
\author{H. Uys}
\email{huys@csir.co.za}
\affiliation{National Laser Centre, Council for Scientific and Industrial Research, Pretoria, South Africa}
\affiliation{Department of Physics, Stellenbosch University, Stellenbosch, South Africa}\author{Humairah Bassa}
%\email{206505721@stu.ukzn.ac.za}
\affiliation{School of Physics, University of KwaZulu, Natal, Durban, South Africa}
\author{Pieter du Toit}
\affiliation{National Metrology Institute of South Africa, Meiring Naude Rd., Pretoria, South Africa}
\author{Sibasish Gosh}
\affiliation{ Institute of Mathematical Sciences, Chennai, India}

\author{T. Konrad}
\email{konradt@ukzn.ac.za}
\affiliation{School of Physics, University of KwaZulu, Natal, Durban, South Africa}
\affiliation{National Institute of Theoretical Physics, UKZN Node}
\begin{abstract}
Measurement combined with feedback that aims to restore a presumed pre-measurement quantum state will yield this state after a few measurement-feedback cycles  even if the actual state of the system initially had no resemblance to the presumed state.  Here we introduce this mechanism of {\it self-fulfilling prophecy} and show  that it can be used to  prepare  finite-dimensional quantum systems in target states  or target dynamics. Using two-level systems as an example we demonstrate that self-fulfilling prophecy protects the system against noise and tolerates imprecision of feedback up to the level of the measurement strength. By means of unsharp measurements the system can be driven deterministically into arbitrary, smooth quantum trajectories. %We hereby give a new operational meaning to the notion of unsharp (weak) measurements.   
\end{abstract}

\maketitle
\section{Introduction.}

The control of individual quantum systems, for example of trapped atoms and ions or single photons, enabled experimental tests of quantum theory and its foundations as well as the application of quantum effects for information processing, communication and  metrology purposes.   The monitoring of observables based on continuous or sequential unsharp (sometimes called weak) measurement \cite{Diosi1988, Belavkin1989, Wiseman1993, Carmichael93, Korotkov2001, Audretsch2001} has paved the way for quantum control in real-time  with closed-loop feedback.  This kind of control already has been applied to photons in  microwave cavities \cite{Sayrin2011} and superconducting qubits \cite{Vijay2012}.  

Here we introduce a new control scheme called self-fulfilling prophecy (SFP), which is related to quantum state monitoring \cite{Doherty.et.al00, Diosi2006, Oxtoby2008, KonradUys2012}.  Both schemes are based on the convergence of different states to a common state subject to sequential measurements  with the same measurement results.  SFP allows to prepare quantum systems in a target state and protect it against decoherence in the presence of  noise and feedback errors. Moreover, it can be employed to drive the system into target dynamics and protect these dynamics. 

The SFP technique uses unitary feedback to return the system into a particular pre-measurement state. The pre-measurement state can also be restored  probabilistically by means of filters or additional measurements.  Such measurement reversals have been used to suppress decoherence \cite{KoashiUeda1999, KorotkovKeane2010, LeeJeong2011} or to protect entanglement \cite{ManXia2012, KimLee2012}.

In what follows, we first revise the formalism for measurement and feedback and describe the protocol of SFP. Then we address the question which measurements are  needed for SFP for systems with finite-dimensional Hilbert space and prove the convergence to the target state for the ideal case without noise or feedback errors.  By means of numerical simulations we study the asymptotic fidelity  in the presence of noise and imperfect feedback for two-level systems. In addition, we employ SFP to protect Rabi oscillation against noise.  We close with two examples of driving two-level systems into target dynamics -- a figure of eight on the Bloch sphere and accelerated Rabi oscillations. 

\section{Quantum measurements  }\label{Section:fidelity}

\begin{figure*}[t]
\centering
\includegraphics[scale=0.5, keepaspectratio]{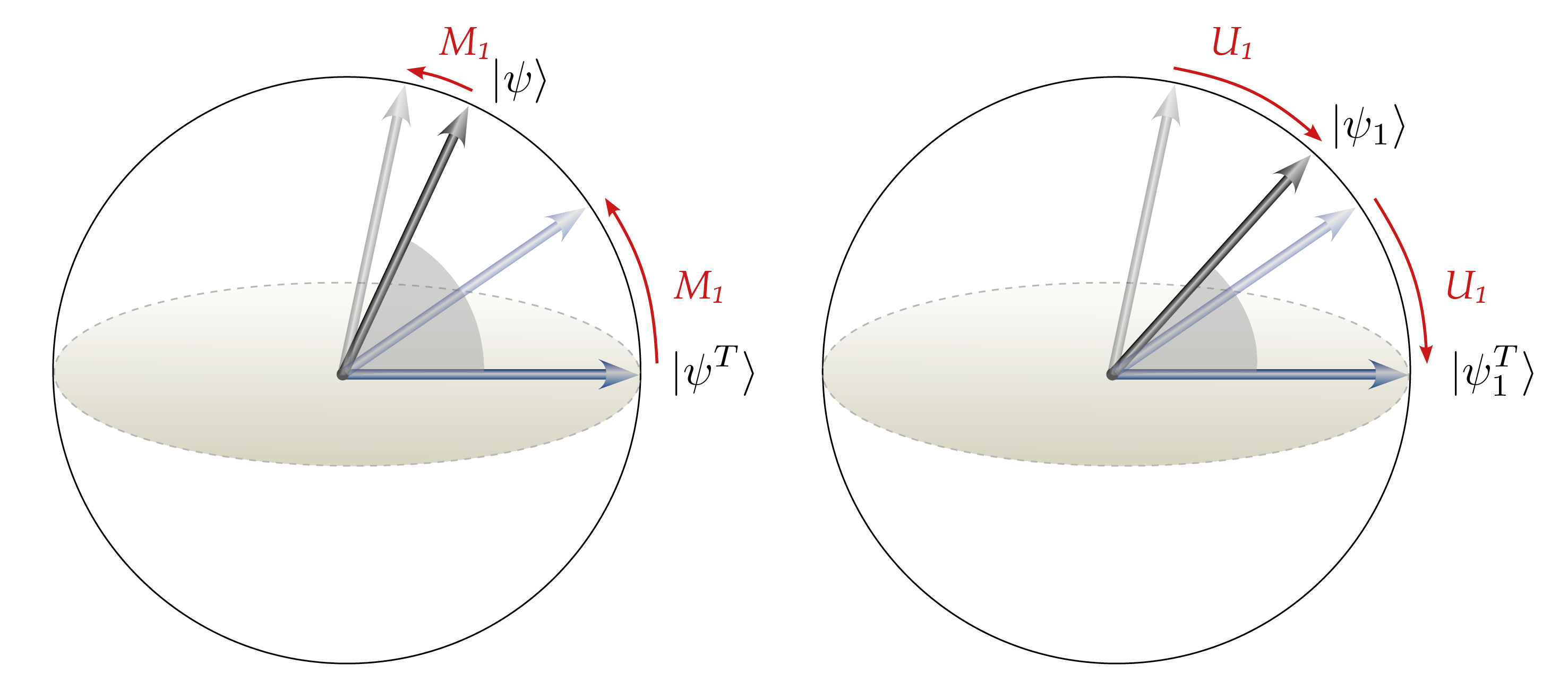}
\caption{In this example of Self-fulfilling Prophecy for a two-level system,  the most probable outcome of an unsharp measurement of the pseudo spin z-component  $\sigma_z$   drives the  state $\ket{\psi}$  towards the north pole of the Bloch sphere (left diagram), which represents the closest eigenstate of the observable.   The unitary feedback $U_1$ (right diagram) is chosen to compensate the back action $M_1$  of the measurement and return the system into its pre-measurement state under the assumption that this was the target state $\ket{\psi^T}$. Afterwards, the system's state $\ket{\psi_1}$  is closer to the target state $\ket{\psi^T_1}=\ket{\psi^T}$ (shaded angle in the right diagram) than before (shaded angle in the left diagram).  \label{Blochsphere}}
%In this example an imagined target qubit state  is aligned along an axis in the equatorial plane in the Bloch picture as shown in (a).  Assume the actual qubit state points upward close to the $z$-axis. Since the actual state is close to an eigenstate of a measurement operator, the measurement has a smaller effect on it compared to the effect on the target state.  If the outcome $M_1$ is observed (for $p_0<1$), then both states are shifted toward the eigenstate $|\!\uparrow\rangle$ as in (b).  The feedback reversal is now applied to both the target state and actual state which returns the target to its original state, and moves the actual state closer to the target  than it had originally been as in (c).  Naturally, had the outcome $M_0$ been observed, the actual state would be left further away from the target at the end of the sequence compared to before.  However, the most probable outcome, as determined by the actual state, is $M_1$.  On average therefore the actual state will move closer to the target.}}\label{Blochsphere}
\end{figure*}

The statistics of measurements in quantum mechanics can be described by means of positive operators $E_i$, so-called {\it effects}, whose expectation values determine the probabilities for the measurement results numbered by the index $i= 1, 2, \ldots$. 
\begin{equation}
p_i(\psi)= \bra{\psi}E_i\ket{\psi}\,.
\end{equation}
 Since the probabilities sum to unity for any state $\ket{\psi}$, the effects sum to the identity operator, $\sum E_i = \mathbb{I}$,  and generate a so-called positive-operator valued measure (POVM). On the other hand, the state of the system after the measurement in general depends on the measurement result $i$ and can be expressed  by so-called Kraus operators $M_i$:
\begin{equation}
\ket{\psi}\xrightarrow{\mbox{result}\, i}\ket{\psi_i}\equiv \frac{M_i}{ \sqrt{p_i}}\ket{\psi}\,.
\end{equation}
The Kraus operators can be decomposed like complex numbers into {\it phase}  and {\it modulus}:
\begin{equation}
M_i = U_i |M_i|\,,
\end{equation} 
where the {\it phase} operator $U_i$ is unitary and can be interpreted as measurement-outcome dependent feedback \cite{Wiseman1993}, while the modulus is related to the effect via $|M_i|\equiv\sqrt{M_i^\dagger M_i}= \sqrt{E_i}$.  An unsharp measurement of a non-degenerate observable $O=\sum_{j=1}^d o_j  \pro{o_j}$ with measurement results $o_1, \ldots, o_d$ is given by a  POVM with commuting effects 
\begin{equation}
E_i=  \sum_{j=1}^d \lambda_{i j}  \pro{o_j}\quad\mbox{where } (\lambda_{ij}) \mbox{ is invertible}, 
\label{nondegenerate}
\end{equation}
such that from the statistics $p_i(\psi)=\bra{\psi}E_i\ket{\psi}$ the probability to measure any $o_j$ can be determined via $|\bra{o_j} \psi\rangle|^2 = \sum_i\lambda^{-1}_{i j} p_i (\psi).$ 
For example, for a two-level system a measurement of the (pseudo-) spin z-component, $\sigma_z=\pro{\uparrow} - \pro{\downarrow}$, is given by the effects:
\begin{eqnarray}
 E_0&=&(1-p_0)|\downarrow\rangle\langle\downarrow| + p_0|\uparrow\rangle\langle\uparrow|\nonumber\\
 E_1&=&p_0|\downarrow\rangle\langle\downarrow| + (1-p_0)|\uparrow\rangle\langle\uparrow|\,, \label{unsharpm}
\end{eqnarray}
Here the squared difference between the eigenvalues  of $E_0$,  $0 < (\Delta p)^2=  (2p_0 - 1)^2   \le 1$,  measures the strength of the measurement with $ (\Delta p)^2 = 0$ a fully weak  and $(\Delta p)^2 = 1$ a strong (von Neumann) measurement of $\sigma_z$ \cite{Audretsch2001, KonradUys2012}.   

\section{Protocol of self-fulfilling prophecy}
Self-fulfilling prophecy transfers  a quantum system with Hilbert space $\mathcal{H}$ of finite dimension $d$ into a target state $\ket{\psi_T}\in \mathcal{H}$ by assuming that the system is initially in the target state $\ket{\psi_T}$ (even though it may not be).  After a measurement unitary feedback is imposed which would return the system into its pre-measurement  $\ket{\psi_T}$, had the assumption been correct. %\TKK{( But even so, the measurement-feedback cycle makes the state of the system approach the target state.}
The condition for the unitary {\it reversal} feedback $U_i$ after a measurement with Kraus operator $|M_i|=\sqrt{E_i}$ thus reads
 \begin{equation}
\ket{\psi^T}\xrightarrow{\mbox{result}\, i} \ket{\psi_i^T}\equiv\frac{U_i \sqrt{E_i}}{\sqrt{w_i}}\ket{\psi^T}= \ket{\psi^T}\,,
\label{feedback}
\end{equation}       
where the normalisation constant is given by $w_i =  \bra{\psi^T}E_i\ket{\psi^T} $. The SFP protocol consists of a number of consecutive executions of the measurement-feedback cycle described above on a system in an unknown state.

\section{Mechanism of convergence}
Executing the SFP  protocol with suitable measurements, the  state of the system comes on average closer to the target state in each measurement-feedback cycle. This is explained graphically for the special case of a qubit measurement in  Fig.\ref{Blochsphere}. 
The proximity (similarity) between the actual state $\ket{\psi}$ and  the target state $\ket{\psi_T}$ can be quantified by the target fidelity, i.e, the squared modulus of the overlap between both states:
\begin{equation}
F(\psi,\psi^T)=\vert\langle{\psi}\ket{\psi^T}\vert^2\,. 
\end{equation}  
%\TKK{$F=1$ means $\ket{\psi_T}=\ket{\psi}$, while $F=0$ indicates that both states can be distinguished with certainty by a single measurement.} 
The change of fidelity $\Delta F$ due to a measurement with feedback averaged over the possible measurement results amounts to
\begin{align}\label{eq:DeltaF}
    \Delta F &=  \sum_i p_i \vert\bra{\psi_i}\psi^T_i\rangle\vert^2~-~\vert\ip{\psi}{\psi^T}\vert^2.\nonumber \\
    & =  \sum_i \frac{\vert\bra{\psi}E_i\ket{\psi^T}\vert^2}{w_i}~-~\vert\ip{\psi}{\psi^T}\vert^2.
\end{align}
%\TKK{In general, any} measurement  brings an arbitrary pair of states $\ket{\psi}, \ket{\psi^T}\in \mathcal{H}$ conditioned on the same outcome $i$ on average closer together or keeps the fidelity the same (monotonicity of \TKK{the average fidelity} of  selective operations) \cite{Rouchon}. 
In general, any measurement carried out on two equal systems with the same measurement result brings an arbitrary pair of states $\ket{\psi}, \ket{\psi^T}\in \mathcal{H}$ on average closer together or keeps the fidelity the same (monotonicity of the average fidelity of  selective operations). This can be seen by rewriting the average change of fidelity  $\Delta F$, and observing that it is positive or zero: 
\begin{align}\label{eq:DeltaF2}
    \Delta F= \sum_i \frac{\vert\bra{\psi}(\mathbb{I} - \pro{\psi^T}) E_i\ket{\psi^T}\vert^2}{w_i} \ge 0\,.
\end{align}    

For SFP, we choose the measurements such that the average fidelity change due to measurement combined with feedback is strictly positive unless the system is in the target state, i.e.,  $\ket{\psi} = \ket{\psi^T}$. This implies  that  on average the fidelity between the state of the system and the target state grows due to the sequence of  measurements until the system reaches the target state.  Since the target state is invariant under the action of  SFP, the system remains in the target state subsequently.

Here we show that there are different kinds of measurement that lead to $\ket{\psi} = \ket{\psi^T}$ (i.e. $F=1$) being  a necessary and sufficient condition for  $\Delta F=0$. For this purpose,  we express the state of the system without restriction of generality as  $\ket{\psi}=\alpha\ket{\psi^T} + \beta\ket{\psi^R}$, where  $\ket{\psi^R}\in \mathcal{H}$ is orthogonal to the target state $\ket{\psi^T}$. It follows that $\alpha=\bra{\psi^T}\psi\rangle$. Moreover, 
\begin{align}\label{eq:DeltaF3}
  & \Delta F= 0 \Leftrightarrow \sum_{i\in\Omega} \frac{\vert\bra{\psi}(\mathbb{I} - \pro{\psi^T}) E_i \ket{\psi^T}\vert^2}{w_i} = 0.
\end{align}
Since each summand in the last equation is greater or equal to zero, all summands must vanish. This is the case if,  and only if, 
\begin{align}\label{eq:DeltaF3}
    &\bra{\psi}E_i \ket{\psi^T} = w_i\bra{\psi}\psi^T\rangle\,\,\mbox{ for all}\,\, i\in\Omega\,.\nonumber \\
    & \Leftrightarrow  (\alpha^*\bra{\psi^T} + \beta^*\bra{\psi^R})E_i\ket{\psi^T} = w_i\bra{\psi}\psi^T\rangle\,\mbox{ for all}\, i\in\Omega\,.\nonumber \\
   & \Leftrightarrow \quad  \beta^*\bra{\psi^R}E_i\ket{\psi^T} = 0 \,\,\mbox{for all}\,\, i\in\Omega\,.
\end{align}   
Hence,  $\beta^*=0$ and thus $|\alpha|^2\equiv F=1$ if,  and only if,  the vectors $E_i\ket{\psi^T}$ span the Hilbert space $\mathcal{H}$ of the system, i.e., for all states $\ket{\psi^R}\in{\mathcal H}$  there is a measurement result $i$  such that  $\bra{\psi^R}E_i\ket{\psi^T}\not= 0$. Thus we found a criterion for  SFP  to drive a quantum system in the absence of noise into the target state. %The measurement effects $E_i$ and the target state $\ket{\psi^T}$ have to be chosen such that}
%\begin{equation}
%\mbox{span}\left\{E_i\ket{\psi^T}\,| i\in\Omega\right\} = \mathcal{H}. 
%\end{equation}

Accordingly, SFP works for any target state $\ket{\psi^T}\in \mathcal{H}$  with informationally-complete measurements, which possess effects $E_i$ that span the space of linear operators on $\mathcal{H}$. This follows from 
\begin{align}
  & 0=\tr{\ket{\psi^T}\bra{\phi} E_i}= \bra{\phi}E_i\ket{\psi^T}\, \mbox{ for all}\, i\in\Omega\,\nonumber\\ & \Rightarrow \ket{\phi}=0\,.
   %& \Rightarrow \quad  \beta^*\bra{\psi^R}E_i\ket{\psi^T} = 0 \,\,\mbox{for all}\,\, i\in\Omega\,.\nonumber 
  \end{align}

\begin{figure}
\centering
\includegraphics[scale=0.5, keepaspectratio]{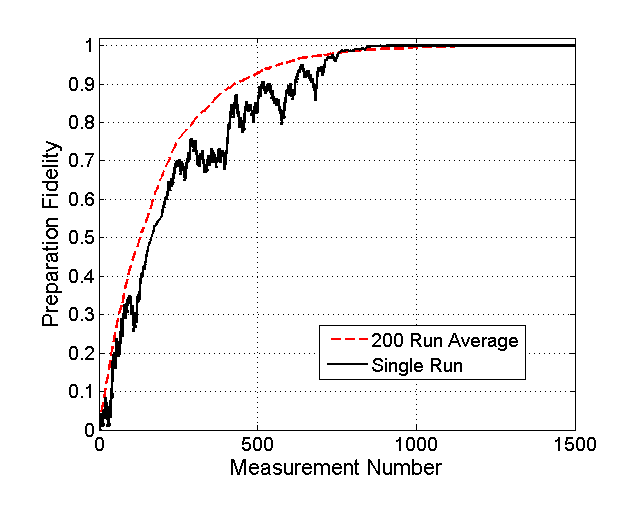}
\caption{Demonstration of convergence of state preparation fidelity.  Even though the actual initial state of the system was orthogonal to the target state, an asymptotic fidelity $F=1$ is reached. Here the dashed red line gives the average over 200 runs, while the black run shows the evolution of the fidelity in a single run.}\label{StatePrepFig}
\end{figure}

%This is the case for informationally complete measurements \TKK{\cite{BuschGrabowskiLahti95}}. The effects $E_i$ of  such measurements form an operator basis on $\mathcal{H}$, and therefore $\tr{E_i A} = 0$  for all $E_i$ implies A = 0. For the special case, $A=\ket{\varphi}\bra{\psi^T}$,  this means  that $\bra{\varphi} E_i\ket{\psi^T}=0$ for all $E_i$ implies $\ket{\varphi}=0$, which proves the claim that the $E_i\ket{\psi^T}$ form a basis of $\mathcal{H}$.

Another important kind of measurement suitable for SFP are unsharp measurements of a non-degenerate observable. For such measurements with results $i=1,\ldots d$ the vectors $E_i\ket{\psi^T}$ form a basis if, and only if, 
\begin{align}\label{get}
&0 \not=\det(E_1\ket{\psi^T}, \ldots, E_d\ket{\psi^T})=\det(( \lambda_{ij}\langle o_j \ket{\psi^T})) \nonumber\\
&\Leftrightarrow 0\not= \det (\Psi \lambda) = \det(\Psi) \det (\lambda)\,.
\end{align}
%where $( \lambda_{ij}\langle o_j \ket{\psi^T})$ is the matrix with the elements $\lambda_{ij}\langle o_j \ket{\psi^T}$.
 This means that neither the determinants of  $\Psi\equiv\sum_{l}  \langle o_l \ket{\psi^T} \pro{l}$ nor the determinant of   $\lambda\equiv \sum_{ij} \lambda_{ij}\ket{j}\bra{i}$  must  vanish for SFP to work.  This condition requires the target state to be a superposition of all eigenstates $\ket{o_l}$ of the measured non-degenerate observable.  Note, that $\det (\lambda)\not= 0$ is satisfied for sharp and unsharp measurements of non-degenerate observables (Eq.~(\ref{nondegenerate})).

We now study the performance of the SFP protocol numerically for qubit control with unsharp measurements. Figure~\ref{StatePrepFig} shows convergence of the state preparation fidelity as a function of the number of measurement- and feedback steps taken.  In this case we chose the target state to be $|\psi^T\rangle = e^{-i\frac{\pi}{8}\hat\sigma_x}|\!\!\downarrow\rangle$, and the initial actual state orthogonal to it.  The solid black line represents the fidelity in a single run of the simulation, while the dashed red line is the average over 200 runs.   The individual measurement strength was $p_0=0.45$. It is conventional to define the strength of a sequence of measurements as $\gamma=\Delta p^2/\tau$, where $\tau$ is the measurement periodicity.  Thus in this case $\gamma=0.01$ in units of the inverse measurement periodicity.  Asymptotically a fidelity of $F=1$ is clearly reached demonstrating that successful state preparation was achieved.

\begin{figure}
\centering
\includegraphics[scale=0.5, keepaspectratio]{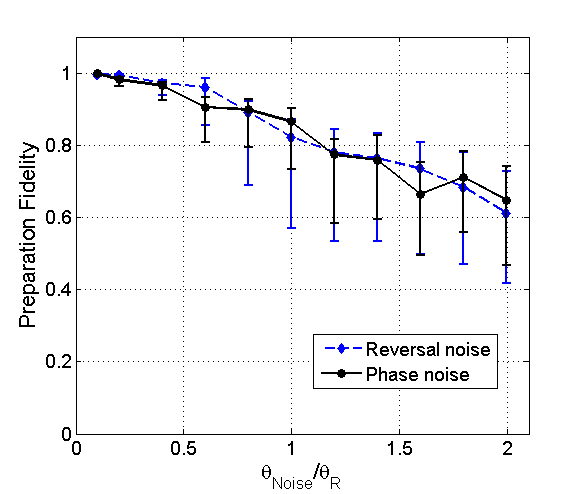}
\caption{Comparison of loss of asymptotic state preparation fidelity due to dephasing noise (black diamonds)  or noise in the reversal operations (blue circles).  In both cases a white noise spectrum was assumed.}\label{NoiseFig}
\end{figure}

In this first example we assumed the absence of any external noise influences.  Now we test the behaviour of the asymptotic fidelity both under the influence of dephasing noise and imperfections in the measurement reversal feedback angle. In both cases we assume that the noise obeys a white noise spectrum and we characterize the strength of the noise by comparing the root-mean-square angular deviation, $\theta_{Noise}$, that the noise causes between successive measurements, to the measurement reversal angle,  $\theta_R := \arccos (\mathbb{R}\textnormal{e}(\bra{\psi^T}U_{i}^\dagger\ket{\psi^T}))=\arccos(\mathbb{R}\textnormal{e}(\bra{\psi^T}\frac{ \sqrt{E_i}}{\sqrt{w_i}}\ket{\psi^T}))$, where $\mathbb{R}\textnormal{e}$ indicates the real part.  In Fig.~\ref{NoiseFig} we plot the asymptotic fidelity by averaging over 6000 measurement and feedback operations in a state preparation run, having used the target state $|\psi^T\rangle = (|\!\!\downarrow\rangle + i|\!\!\uparrow\rangle/\sqrt{2})$. Above 90\%  state preparation fidelity can be achieved as long as $\theta_{Noise}\lesssim\theta_R/2$ for both dephasing noise, circles, and noise in the reversal, diamonds.  The error bars indicate the root-mean-square deviations above and below the mean.  This demonstrates that the feedback scheme can preserve qubit states  asymptotically long with high fidelity while tolerating modest noise influences. The scheme does require that the target qubit state is known.

\begin{figure}
\centering
\includegraphics[scale=0.5, keepaspectratio]{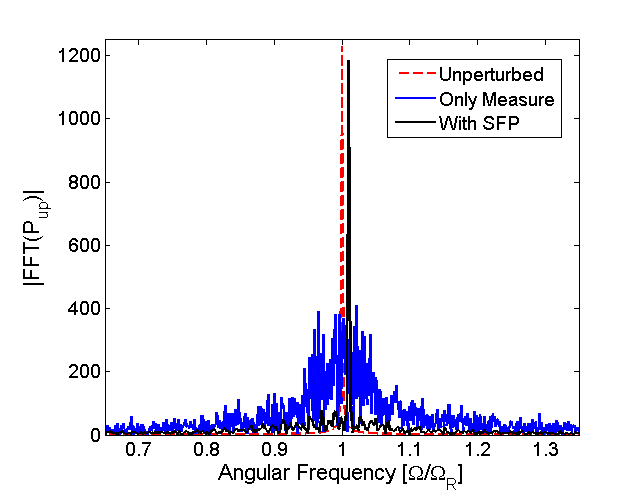}
\caption{SFP forces a qubit oscillating at frequency $\Omega_R=1$ (red, dashed durve) to oscillate instead at a target frequency $\Omega_T=1.01$ (black curve).  Measurement without reversal allows state estimation \cite{KonradUys2012}, but also leads to significant broadening of the oscillation spectrum (blue curve).}\label{FreqShiftFig}
\end{figure}

Now we employ the measurement reversal protocol to influence a separate underlying unitary dynamics. In particular we study a qubit undergoing Rabi oscillations at an angular frequency of $\Omega_R=1.00$.  We desire that the qubit oscillates instead at a target angular frequency of $\Omega_T=1.01$.  In addition, a target  state is taken to initially be orthogonal to the initial actual state.  The actual- and target states are then time evolved by Rabi oscillations at the actual- and target frequencies respectively. To control the actual frequency a self-fulfilling prophecy approach is again used.  We simply assume that the actual state is undergoing the dynamics of the target state. A sequence of unsharp measurements are made on the actual state, but each time the measurement is reversed by assuming that it is instead in the target state as predicted by the target dynamics.  In Fig.~\ref{FreqShiftFig} we plot the discrete Fourier transform of the probability for the qubit to be in the upper state.  Without measurement and reversal the dashed red line indicates that the system is oscillating at the Rabi frequency $\Omega_R=1$.  The blue curve indicates that attempts to monitor the qubit using unsharp measurements induce significant measurement noise, leading to significant broadening of the oscillation spectrum. Once the  reversals are initiated the noise is suppressed and the frequency shifts to the target frequency.  This approach works as long as the sequential measurement strength, $\gamma$, is larger than the frequency detuning $\delta=\Omega_R-\Omega_T$.
\begin{figure}
\centering
\includegraphics[scale=0.5, keepaspectratio]{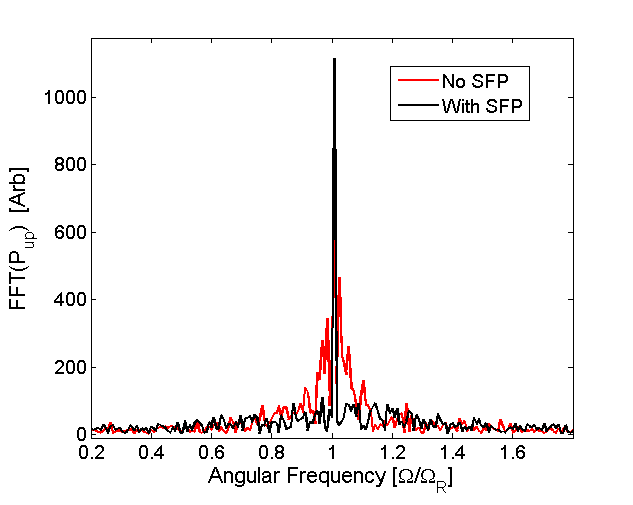}
\caption{Suppression of noise using SFP.  The red curve shows the spectrum of a qubit oscillating in the presence of white noise on the drive field amplitude.  The black curve shows the stabilized spectrum. }\label{StabilizationFig}
\end{figure}

SFP can also be used to suppress noise present in the unitary dynamics.  In Fig.~\ref{StabilizationFig} we show the noise spectrum of a qubit oscillating in the presence of white noise on the amplitude of the drive field. The root-mean white noise field amplitude was $1/2$ the drive field strangth, thus leading to the broadened spectrum (red curve).  Implementing SFP with $\gamma=0.4$ clearly leads to a strong suppression of the noise. We used 40 measurements per Rabi oscillation cycle and $p_0=0.45$.

\begin{figure}
\centering
\includegraphics[scale=0.5, keepaspectratio]{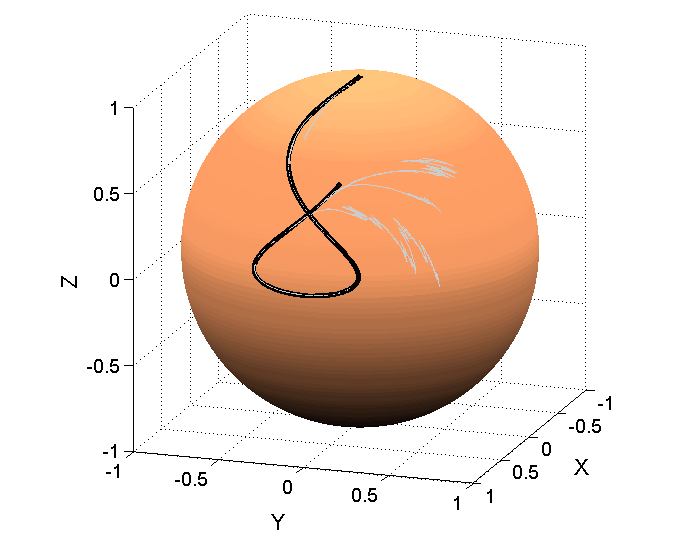}
\caption{A target dynamics forming a figure of eight on the Bloch sphere is imagined.  SFP forces a qubit to execute this dynamics.  Here the different gray curves are due to different choices for the actual initial state. Each curve quickly converges onto the target dynamics.}\label{TrajectoryFig}
\end{figure}

Finally, we show that the state preparation scheme can be adapted to elicit qubit dynamics on its own, without an additional unitary dynamics. To this end we imagine the target state to change dynamically and adapt the feedback reversal to the instantaneous target state, but still execute the same unsharp measurement. As long as the imagined dynamics is slow compared to the timescale of convergence the actual state will follow the target dynamics.  This expectation is clearly borne out in Fig.~\ref{TrajectoryFig}. 

Here we chose a figure-of-eight trajectory on the Bloch sphere for the target dynamics and three different starting points for the actual state.  Over the trajectory completion time 10 000 measurements were executed using $p_0=0.45$ as the strength of individual meausurements. In units of the trajectory completion time $\gamma=100$, indicating that the dynamics is strongly dominated by the measurement and feedback.   From all three starting points the actual state quickly converges to the target state and then dynamically follows it.  In the continuous measurement limit this constitutes a new class of measurement and feedback driven qubit dynamics.  Unlike pure unitary evolution the dynamics can be preserved in the long time limit even in the presence of modest noisy influences.

The work in this paper was supported in part by the National Research Foundation of South Africa through grant no.~93602 as well as an award by the United States Airforce Office of Scientific Research, award no.~ FA9550-14-1-0151.

\bibliographystyle{unsrt}
\bibliography{xbib}

\end{document}